\documentclass[11pt,twoside]{article} 
\usepackage{asp2004}
\usepackage{epsf}
\usepackage{psfig}
\usepackage{lscape} 

\begin{document} 
\title{NLTE Analysis of the Ultra-short Period White-dwarf Binary 
RX\,J0806.3+1527}
\author{J. Steiper, K. Reinsch, and S. Dreizler} 
\affil{Institut f\"ur Astrophysik, Universit\"at G\"ottingen, 
Friedrich-Hund-Platz 1, D-37077 G\"ottingen, Germany}

\begin{abstract} 
RX\,J0806.3+1527 is suspected to be a double-degenerate white dwarf
binary.  We present first results of our NLTE analysis of its optical
spectrum. The VLT/FORS1 data show a composite spectrum consisting of a
blue continuum and superimposed emission lines of the He\,{\sc ii}
Pickering series and, possibly, the H Balmer series. Our models are
based on hot white dwarf atmospheres and include illumination effects
onto the secondary star. The physical parameters and chemical
abundances derived from the comparison of the observed spectrum with a
grid of model atmospheres provide constraints on the true nature of
this enigmatic binary and on the models proposed so far.
\end{abstract}

\section{Introduction}
RX\,J0806.3+1527 is a remarkable object discovered during the ROSAT all-sky 
survey \citep{beuetal99}. Its soft X-ray flux is $\approx$ 100\% modulated with 
a period of 321.54\,s \citep{isretal99,burrei01}. If this is the orbital period 
of a double-degenerate close binary system as suggested by \citet{isretal02} 
and \citet{rametal02} it would lie close to the theoretical minimum period of 
white dwarf binaries and would imply the presence of a He-rich donor star.

The spectral energy distributions in the optical and X-ray windows reveal two
distinct components with black body temperatures of $\approx$\,40\,000\,K and
$\approx$\,750\,000\,K which could be attributed to the donor star or an
accretion disk and to an accretion spot on the primary, respectively 
\citep{isretal03}.
As a crucial test to distinguish between the double-degenerate scenario and 
competing models \citep[e.g.][]{noretal04} for RX\,J0806.3+1527 we have started 
an abundance analysis of the optical spectrum presented by \citet{isretal02}.

\section{Spectral Analysis of RX\,J0806.3+1527}

The optical spectrum of RX\,J0806.3+1527 obtained with FORS1 at the ESO VLT
shows a composite of a blue continuum and superimposed emission lines of the 
He\,{\sc ii} Pickering series and, possibly, the H Balmer series.
For the spectral analysis, we utilize a grid of theoretical spectra calculated 
with an NLTE model atmosphere code for white dwarf photospheres with a range
of effective temperatures $T_{\rm wd}$, $(He/H)$ abundance ratios, and 
isotropic irradiation \citep{werdre99,dre03,weretal03}. Here, we present a 
study of the influence of the three main parameters on the spectrum.

\subsection{Illumination Parameter}

\begin{figure}[!ht]
  \plotone{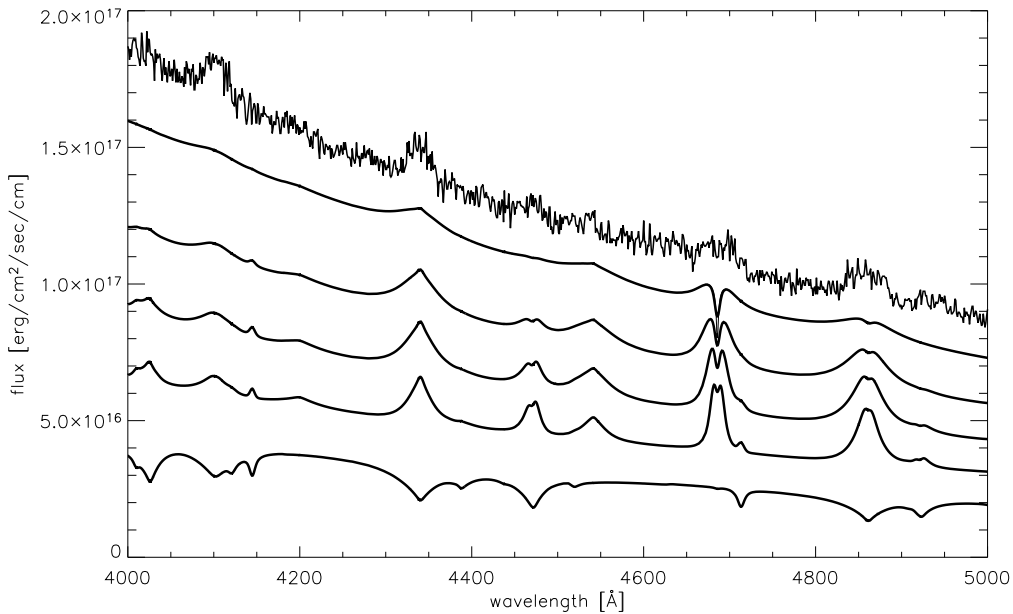}
  \caption{Observed spectrum of RX\,J0806.3+1527 in comparison to a selection 
  of models with different illumination parameters $i$ (from top to bottom:
  observation, models with $i = 0.0002, 0.0001, 0.00005, 0.00002$, and model
  without irradiation. For comparison with the models, the observed spectrum 
  has been arbitrarily offset in flux units.}
  \label{fig:illumination_parameter}
\end{figure}

In our models, irradiation is treated as black body emission with
effective temperature $T_{\rm irr} = 500\,000$\,K from a spherically
symmetric shell around a white dwarf with $T_{\rm wd} = 30\,000$\,K
and an abundance ratio $(He/H) = 1.0$. The distance between the
irradiating source and the photosphere is represented by a free
``illumination'' parameter $i$. It has been varied from $10^{-1}$ to
$10^{-10}$ with a more detailed study in the parameter range which
leads to the occurrence of significant emission lines in the spectrum
(Fig.~\ref{fig:illumination_parameter}). Depending on $i$, our
models produce three phenomenologically different types of spectra:

\medskip \noindent
\begin{tabular}{@{}l@{\ }c@{\ }c@{\ }c@{\ }l@{:~}p{9.5cm}}
  $10^{-10}$ & $<$ & $i$ & $<$ & $10^{-5}$ & 
  Absorption lines are gradually filled by emission \\
  $2\!\cdot\!10^{-5}$ & $<$ & $i$ & $<$ & $2\!\cdot\!10^{-4}$ &
  Emission lines with central absorption in He\,{\sc ii} $\lambda$ 4686\,\AA\ \\
  $2\!\cdot\!10^{-4}$ & $<$ & $i$ & & & 
  Featureless spectrum except for He\,{\sc ii} $\lambda$ 4686\,\AA\ absorption 
  \\
\end{tabular}

\subsection{\boldmath $(He/H)$ Abundance Ratio}

\begin{figure}[!ht]
  \plotone{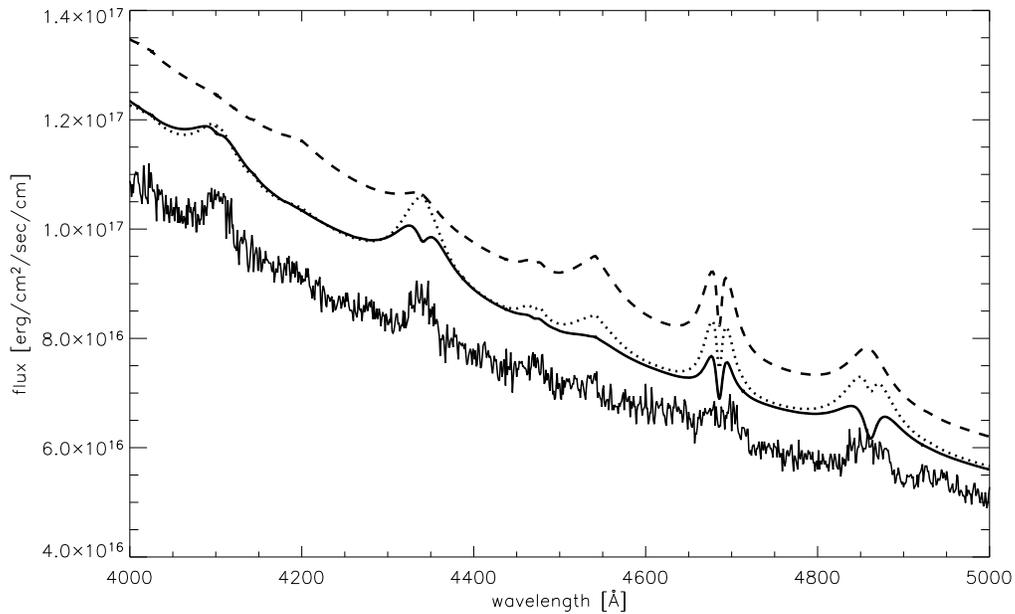}
  \caption{Sequence of models with different $(He/H)$ abundance ratios (from
  top to bottom: $(He/H) = 100, 0.3, 0.1$) and observed spectrum of 
  RX\,J0806.3+1527 (lowermost curve). For comparison with the models, the 
  observed spectrum has been arbitrarily offset in flux units.}
  \label{fig:abundance_ratio}
\end{figure}

In order to estimate the chemical composition of RX\,J0806.3+1527 we have
calculated models with different $(He/H)$ abundance ratios between $(He/H) =
0.01$ and $(He/H) = 100$. The irradiation and photospheric temperatures have
been fixed at $T_{\rm irr} = 500\,000$\,K and $T_{\rm wd} = 30\,000$\,K,
respectively, and the illumination parameter has been set to $i = 0.0001$ 
(see Fig.~\ref{fig:abundance_ratio}). 

Based on our analysis, we can reject He-deficient atmospheres, $(He/H) \le 
0.01$, as well as models with strong He overabundances, $(He/H) \ge 1.0$. 
The latter fail to reproduce the observed spectrum, especially around the 
He\,{\sc ii} $\lambda$ 4540\,\AA\ line. Proper fits to the observational data 
can be obtained with abundances $0.1 \le (He/H) \le 0.3$.	  

\subsection{White Dwarf Temperature}

\begin{figure}[!ht]
  \plotone{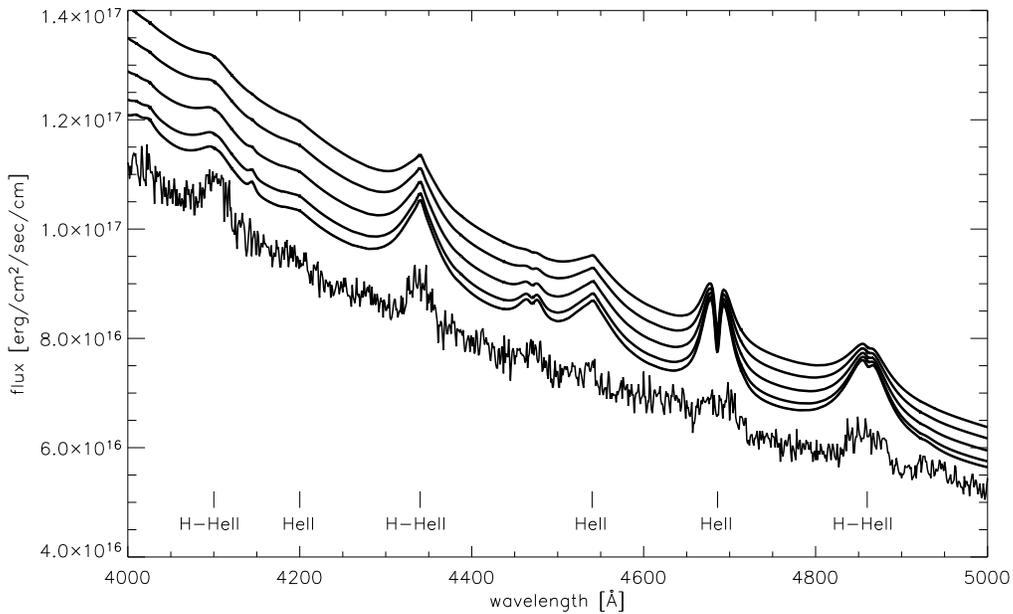}
  \caption{Sequence of models with different white dwarf temperatures 
  $T_{\rm wd}$ (from top to bottom: $T_{\rm wd} = 50\,000$\,K to $30\,000$\,K 
  in steps of $5\,000$\,K) and observed spectrum of RX\,J0806.3+1527 
  (lowermost curve). For comparison with the models, the observed spectrum 
  has been arbitrarily offset in flux units.}
  \label{fig:effective_temperature}
\end{figure}

As a third parameter, we varied the effective temperature of the illuminated
photosphere in the range $30\,000$\,K $\le T_{\rm wd} \le 50\,000$\,K with 
fixed parameters $i = 0.0001$ and $(He/H) = 1.0$ (see Figure 
\ref{fig:effective_temperature}).

Our models show that photospheric temperatures between 30\,000\,K and 
35\,000\,K seem to be more realistic than higher temperatures, in agreement 
with the overall spectral energy distribution of RX\,J0806.3+1527. 

\section{Results and Future Work}

Our parameter study has shown that the optical spectrum of RX\,J0806.3+1527 
can be quantitatively reproduced with our irradiated white dwarf atmosphere 
model. The model spectra provide some evidence against a He-white dwarf nature
of the donor star implied by the double-degenerate binary scenario.
Noting that illumination is the dominant effect in modeling the optical 
spectrum of RX\,J0806.3+1527, our approach of an isotropic irradiation of a
high-gravity atmosphere may, however, be too simplistic. Further work will be 
required to investigate models with non-isotropical irradiation and irradiated 
accretion disks.


\acknowledgements{J. Steiper thanks the organizers of the workshop for 
financial support.}

\end{document}